\documentclass{aa}
\usepackage{graphics}
\usepackage{longtable} 
\usepackage{natbib}
\begin{document}

\title{NEXXUS:  A comprehensive ROSAT survey of coronal X-ray emission among nearby 
solar-like stars \thanks{\sl Tables 1-3 
are also available in electronic form
at the CDS via anonymous ftp to cdsarc.u-strasbg.fr (130.79.128.5)
or via http://cdsweb.u-strasbg.fr/cgi-bin/qcat?J/A+A/
}}
\author{J. H. M. M. Schmitt\and C. Liefke}
\institute{
Hamburger Sternwarte, Universit\" at Hamburg, Gojenbergsweg 112,
D-21029 Hamburg, Germany\\
\email{jschmitt@hs.uni-hamburg.de; cliefke@hs.uni-hamburg.de}}

\titlerunning{NEXXUS}
\offprints{J.H.M.M. Schmitt}
\date{Received 19 February 2003 / Accepted 20 March 2003}

\abstract{We present a final summary of all ROSAT X-ray observations of nearby stars.
All available ROSAT observations with the ROSAT PSPC, HRI and WFC have been matched with
the CNS4 catalog of nearby stars and the results gathered in the {\bf Ne}arby {\bf X}-ray
and {\bf XU}V-emitting {\bf S}tars data base, available via www from the 
Home Page of the Hamburger Sternwarte at the URL 
{\it http://www.hs.uni-hamburg.de/DE/For/Gal/Xgroup/nexxus}.  New volume-limited samples
of F/G-stars ($d_{lim}$ = 14 pc), K-stars ($d_{lim}$ = 12 pc), and M-stars 
($d_{lim}$ = 6 pc)
are constructed within which detection rates of more than 90 \% are obtained; only one star
(GJ~1002) remains undetected in a pointed follow-up observation.  
F/G-stars, K-stars and
M-stars have indistinguishable surface X-ray flux distributions, and the lower envelope
of the observed distribution at $F_X \approx$ 10$^4$ erg/cm$^2$/sec is the X-ray flux level 
observed
in solar coronal holes.  Large amplitude variations in X-ray flux are uncommon for 
solar-like stars,  but maybe more common for stars near the bottom of the main sequence;
a large amplitude flare is reported for the M star LHS 288.  Long term X-ray light curves
are presented for $\alpha$ Cen A/B and Gl~86, showing variations on time scales of weeks 
and demonstrating that $\alpha$ Cen B is a flare star.

\keywords{stars: activity -- stars: coronae -- stars: late type -- X-rays: stars}
}

\maketitle

\section{Introduction}

The discovery of X-ray emission from normal stars with the {\it Einstein Observatory} 
\citep{Vaiana81} and ROSAT demonstrated the ubiquity of hot coronae around late-type stars. 
Specifically, the volume-limited surveys by \citet{Schmitt95} and
\citet{Schmitt97} showed that coronal formation is universal 
in the sense that coronae are formed basically for all (or almost all) main-sequence stars 
with outer convection zones.  Furthermore, the state of ``minimum energy'' of a stellar 
corona (around a main sequence star) was found to be one whose mean X-ray surface flux 
corresponds to the surface flux of a solar coronal hole.  \citet{Hunsch99} and
\citet{Sterzig97} 
extended those studies to all stars within 25 pc by cross-correlating the 
ROSAT all-sky survey (RASS) data with entries in in the Gliese catalog (CNS3) of nearby 
stars and detected a significant fraction of stars listed in CNS3 as X-ray sources 
in the all-sky survey data, while \citet{Hunsch98b} and \citet{Hunsch98} carried out similar
studies for bright main sequence and giant stars in the Bright Star Catalog and
\citet{schr98} presented a volume-limited sample of giant stars within 35 pc around
the Sun.  Similar studies on early-type stars were presented by \citet{berg97}, who 
present a catalog of bright O- and B-type stars detected in the ROSAT all-sky survey, while
\citet{Flem96} discuss X-ray emitting white dwarfs detected in the ROSAT all-sky survey.
The dearth of X-ray emission among A-type stars has already been noted on the basis
of observations with the {\it Einstein} Observatory \citep{Schm85}, \citet{Hunsch01} presents a similar
study from the ROSAT all-sky survey data, and \citet{Simon95} from pointed ROSAT data. 
As to open clusters, most of them are too distant for sensitive studies with ROSAT survey
data, so most of the ROSAT work on open clusters was carried using pointing data; for an
overview of this work see \citep{Rand00}.  The Hyades cluster, however, has very a large angular extent
because of its proximity and can be comprehensively studied only with survey data.  
\citet{Stern95} present the as of to date most complete X-ray study of the Hyades, while 
\citet{Schm93} present the ROSAT all-sky survey data on the core of the Pleiades
cluster.

One of the key properties of late-type stars is their X-ray variability.  The ROSAT all-sky
survey with its approximately 30 sec snapshot exposures extending over two days and longer
provided a unique sampling pattern of coronal X-ray emission.  In particular, long-duration
flares and rotational modulation can be well studied with such data.  \citet{Schmitt94}
discusses ROSAT survey observations of flare stars, and \citet{Haisch94} study specifically 
the X-ray variability of giants observed during the all-sky survey.
\citet{Fuhr03} present a systematic study of X-ray variability in the RASS data, finding that 
stars are indeed the most variable class of X-ray emitters.

Meanwhile the ROSAT operations have ended and no further ROSAT data will be taken.
Data from numerous individual pointings in particular with the ROSAT HRI 
have become available in the ROSAT archive, 
the ROSAT survey data have been
reprocessed, and the Gliese catalog of nearby stars has been substantially updated.
The purpose of this paper is therefore to revisit and supplement the studies by 
\citet{Schmitt97} and \citet{Sterzig97} by utilizing all available -- and presumably
final since reprocessed -- ROSAT data.  
Specifically, \citet{Schmitt95}, \citet{Schmitt97} and \citet{Sterzig97}
had to use the pre-HIPPARCOS distance scale, but far more accurate HIPPARCOS 
parallaxes are now available for many nearby stars.  Interestingly, the parallaxes for the
brighter stars tended to be ``older'' hence less reliable so that the composition of the
volume-limited samples has changed.  Further, the
ongoing infrared all-sky surveys provide ``new'' stars even in the immediate solar vicinity.
Especially for K-type stars more ROSAT data on nearby stars were taken during the last 
years of its lifetime.  In summary, now appears to be a good opportunity for a definitive 
summary of all ROSAT observations of nearby stars.
 
\section{Observations}
The ROSAT Observatory was operated between 1990 - 1998. Between July 1990 and January 1991
it carried out its ROSAT All-Sky Survey (RASS) with the ROSAT Position Sensitive Proportional
Counter (PSPC). Afterwards  pointed observations of individual X-ray sources were carried out in
the framework of the ROSAT guest investigator program both with the PSPC and a 
High (angular) Resolution Imager (HRI).  These detectors had fields of view of about 
7000 $arcmin^{2}$ for the PSPC and
1000 $arcmin^{2}$ for the HRI, so that many X-ray sources were picked up serendipitously 
in the field of view of many observations whose original scientific goal was actually quite 
different.  A boron filter could be placed in front of the PSPC detector for pointed observations
allowing to separate the X-ray band below 0.28 keV (i.e., the ``carbon'' band) into two
separate energy bands; however, only a rather small number of observations was carried out with 
the boron filter in place.

\subsection{X-ray data}
 
The results of both the RASS observations and the ROSAT pointed observations are available 
in the ROSAT results archive in the form of source lists. All five source catalogs used 
for this study, i.e. the ROSAT Bright Source and Faint Source Catalog, the Second ROSAT 
Source Catalog of Pointed Observations with the PSPC with and without filter and the First 
ROSAT Source Catalog of Pointed Observations with the HRI as well as detailed information
on the detection and screening procedures applied in the construction of the catalogs
are available via www from the 
ROSAT Home Page at Max-Planck-Institut f\"ur Extraterrestrische Physik 
{\it http://wave.xray.mpe.mpg.de/rosat/catalogue} or its mirror sites.
These catalogs were our primary source of information, and only in individual cases (discussed
explicitly in our paper) did we go back to the original X-ray data.  It is important to
realize in this context that the above source catalogs have been constructed with rather 
conservative detection thresholds.  The use of these conservative thresholds was mandatory since 
X-ray sources were searched for everywhere in all the ROSAT images.  Since we are interested 
in X-ray emission only from specific locations (i.e., at the positions of nearby stars), we
are working with a far smaller number of trial positions and can therefore
``afford'' choosing lower acceptance thresholds without compromising on the number of 
spurious sources.   It is therefore possible in principle to obtain X-ray ``detections'' from
stars not listed in the above catalogs and in our data base; specific cases discussed
below are the nearby K star Gl~653 and the nearby M-dwarf LHS 288 . 

\subsection{Optical data}

As our source of optical data we used the CNS4 compilation of nearby stars compiled
by \citet{Jahreiss02}.
As pointed out above, the ongoing ever more sensitive infrared surveys provide ``new'' stars 
in the immediate vicinity of the Sun such as the M9V dwarf DENIS-P J104814.7-395606.1
at a distance of possibly 4.1 $\pm$ 0.6 pc \citep{Delfosse01} or LHS 2090
\citep{Scholz01}, although very often the parallax 
information on those stars is extremely limited.  Therefore no compilation of nearby stars 
can be truly complete. Still, among the F, G and K type stars the CNS4 catalog ought to be 
complete and the only source of error should be the parallax error which can move a given 
star outside or inside a specified sampling volume. On the other hand, among the very late 
M stars and brown dwarfs the CNS4 catalog is bound to be incomplete.

\section{Data Analysis} 

The ROSAT source lists and the CNS4 catalog were searched for positional coincidences. As a
matching criterion we used 120 arcsec for survey data, 60 arcsec for pointing data with
the PSPC and 30 arcsec for HRI data.  The differently chosen positional acceptance thresholds 
reflect the
fact that, first, the intrinsic positional accuracy of survey data, 
and the PSPC and HRI pointing data
decreases in that order, and second, the CNS4 input catalog does not have the most
accurate positional information since it is not intended to be a positional catalog. 
The chosen acceptance thresholds do not introduce significant errors into our 
X-ray source lists.
In quite a few cases multiple detections of coronal X-ray emission from
individual stars are available.
Specifically, many of the nearby stars observed and 
detected made in the ROSAT pointing program were also detected in the survey data.
  
We stress that X-ray source identifications are made solely on the basis of positional 
coincidence.  Nevertheless we expect the number of spurious identifications to be very 
small.  Calculating the number of identifications obtained by distributing approximately
100 000 RASS sources over 3231 positions (i.e., the number of Gliese stars) with a detect cell
radius of 2 arcmin, results in $\sim$ 27 spurious identifications or 2\% of the total number 
of RASS detections of nearby stars.  The actual distribution of position offsets 
is much narrower.
To be specific, only 16 out of 1217 survey detections have position offsets of more 
than 100 arcsec, 
and only 80 X-ray detections are off by more than 60 arcsec from the optical positions.  
We thus
conclude that the fraction of incorrectly identified X-ray emitters is at the
one percent level at worst. 

\section{The NEXXUS data base}

Altogether we can associate 1333 of the 3231 
stars up to a distance of 25 pc contained in the CNS4 catalog with ROSAT detected 
X-ray sources; the vast majority of these sources comes from the RASS data (1217).
The remaining 116 stars were only detected in pointed observations.  In addition to the survey 
detections, 328 stars were also observed with the PSPC in pointed mode without the boron filter, 
49 with the filter, and 242 stars were observed with the ROSAT HRI. Moreover, 
some of these stars were observed more than once, so that multiple detections of 
coronal X-ray emission are available.
For easy access and future reference
the results of this cross-correlation process were assembled in the {\bf Ne}arby {\bf X}-ray
and {\bf XU}V-emitting {\bf S}tars data base, available via www from the 
Home Page of the Hamburger Sternwarte at the URL 
{\it http://www.hs.uni-hamburg.de/DE/For/Gal/Xgroup/nexxus}.  
For each CNS4 star detected in X-rays the database provides detailed information about the star 
itself, e.g. several catalog names, coordinates, proper motion, apparent and absolute magnitude, 
color indices, parallax and distance; and information about the associated X-ray source(s), 
e.g. coordinates, count rates with errors, background, exposure time, and an X-ray luminosity. 
For PSPC observations hardness ratios and a source likelihood are given, for HRI observations a 
signal to noise ratio. The date of observation, an off-axis angle, 
and source and sequence numbers 
of pointed observations are listed, too.  
To facilitate browsing, NEXXUS can be searched by 
coordinates, catalog name, color index, magnitude, proper motion or distance.  For completeness we also list the
XUV measurements of the CNS4 stars obtained with the ROSAT Wide Field Camera (WFC); the WFC data
were kindly made available to NEXXUS by J. Pye (Leicester University).  The WFC survey data is
published by \citet{Pye95}, the pointed WFC-data will be discussed by Pye et al. 
(2003, in preparation). 

\section{Results}

\begin{figure}
\resizebox{\hsize}{!}{\includegraphics{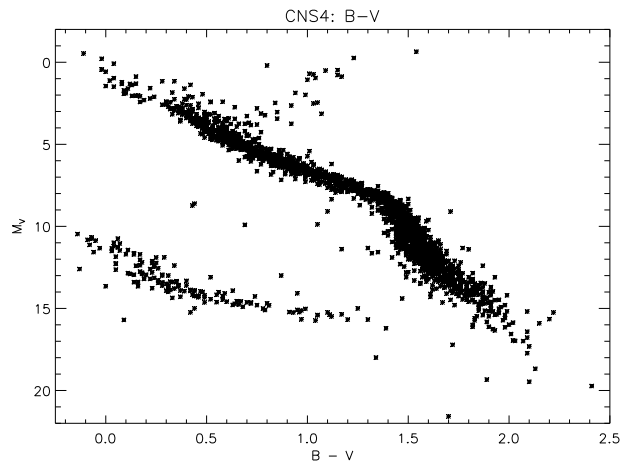}}
\caption[]{\label{CNS4} Color-magnitude diagram ($B-V$ vs. $M_V$) for all stars contained in CNS4 
catalog.}
\end{figure}

\begin{figure}
\resizebox{\hsize}{!}{\includegraphics{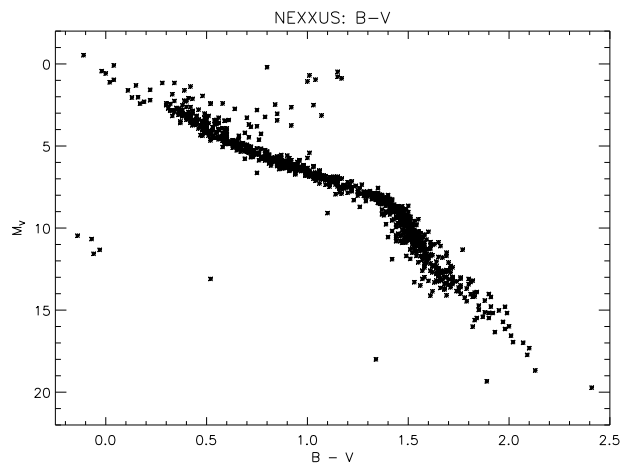}}
\caption[]{\label{NEXXUS}Color-magnitude diagram ($B-V$ vs. $M_V$) for all X-ray detected stars listed
in the NEXXUS data base. }
\end{figure}

From the above cited numbers the overall ROSAT detection rate of the nearby stars listed in CNS4
is 41.1 \%.  It is instructive to compare a color-magnitude diagram of all CNS4 stars 
(cf., Fig. \ref{CNS4})
with a similar diagram constructed only for NEXXUS stars (cf., Fig. \ref{NEXXUS}).   One immediately
notes the well-known paucity of X-ray emitting white dwarfs as well as the absence of X-ray detected
brighter giants.  Stellar X-ray emission is detected down to an absolute 
magnitude of $M_V = $20, i.e., down to the very bottom of the main sequence.


Except for white dwarfs and the brighter giants the overall detection rate of 41.1 \% is due 
to the lack of sufficient sensitivity for the more distant CNS4 stars. In the following we therefore
focus on the very nearest stars for which almost complete detections
are available.  For this purpose we introduced three categories of stars.
Since color information and spectral type are not always consistent, we decided to use a
criterion based on absolute magnitudes to group our sample stars into three categories, 
those with $M_V$ in the range 3 $\le M_V \le $5.80 i.e., the F/G-stars, those 
with  5.80 $< M_V \le$ 8.50, i.e., the K stars, and those fainter
than $M_V$ = 8.50, i.e., the M-type stars. We explicitly excluded all
giants or white dwarfs from these samples; specifically, the giant stars Capella (Gl~194AB),
a well-known RS CVn-like binary, $\beta$ Gem (= Gl~286;~ \citet{Hunsch98}) and 
Arcturus (= Gl~541;~\citet{Ayres91}) are not listed here. 
Tab. \ref{fgstars} contains the results for the F/G-stars, Tab. \ref{kstars} those
for the K stars, and Tab. \ref{mstars} those for stars fainter than $M_V = $8.50. 
In each table we provide the stars' names, their absolute magnitude,
their spectral type and distance as listed in the CNS4 catalog provided
by \citet{Jahreiss02}. For each star we then list the measured count rate and its error,
the exposure time and an X-ray luminosity. 
In order to convert from count rate to X-ray flux we used a conversion factor
of $6 \times 10^{-12}$ erg cm$^{-2}$ ct$^{-1}$ for PSPC data, 3 $\times$ 
10$^{-11}$ {erg} cm$^{-2}$ ct$^{-1}$ 
for PSPC with boron filter and 2.4 $\times$ 10$^{-11}$ erg cm$^{-2}$ ct$^{-1}$ for HRI data; 
X-ray luminosities were compiled from these X-ray fluxes and the HIPPARCOS distances.
For RASS and PSPC pointing data the source likelihood is given, for HRI data the 
signal to noise ratio of the detection. 
A flag indicates the source of the data; S stands for
RASS data, P for PSPC pointing data without the boron filter, F for PSPC pointing data with
the boron filter, and H for HRI data. If a star was observed more than once with the same instrumental setup, only the observation with the longest exposure time is listed.
In the case of binaries the components are listed
separately if they were resolved in any of the ROSAT observations, only one entry is given for
binaries unresolved by ROSAT.
As far as the X-ray data are concerned, Tab. 1-3 are meant
to replace Tab. 1 in \citet{Schmitt95} and Tab. 2 in \citet{Schmitt97}.

\subsection {Detection completeness}

\begin{figure}
\resizebox{\hsize}{!}{\includegraphics{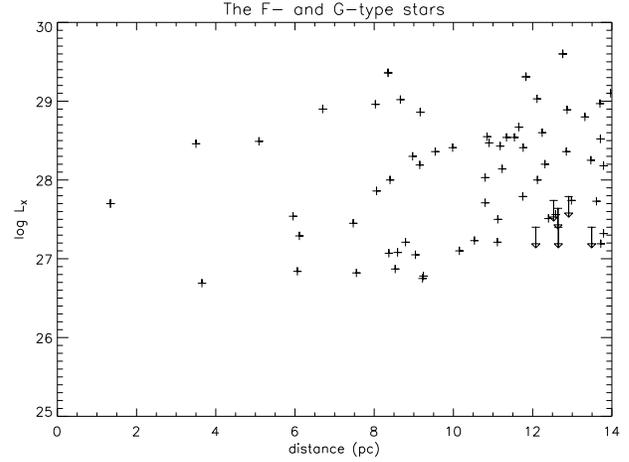}}
\caption[]{\label{FG} $L_X$ vs. distance for the F/G star sample; note the
complete sampling out to 12 pc.}
\end{figure}
\begin{figure}
\resizebox{\hsize}{!}{\includegraphics{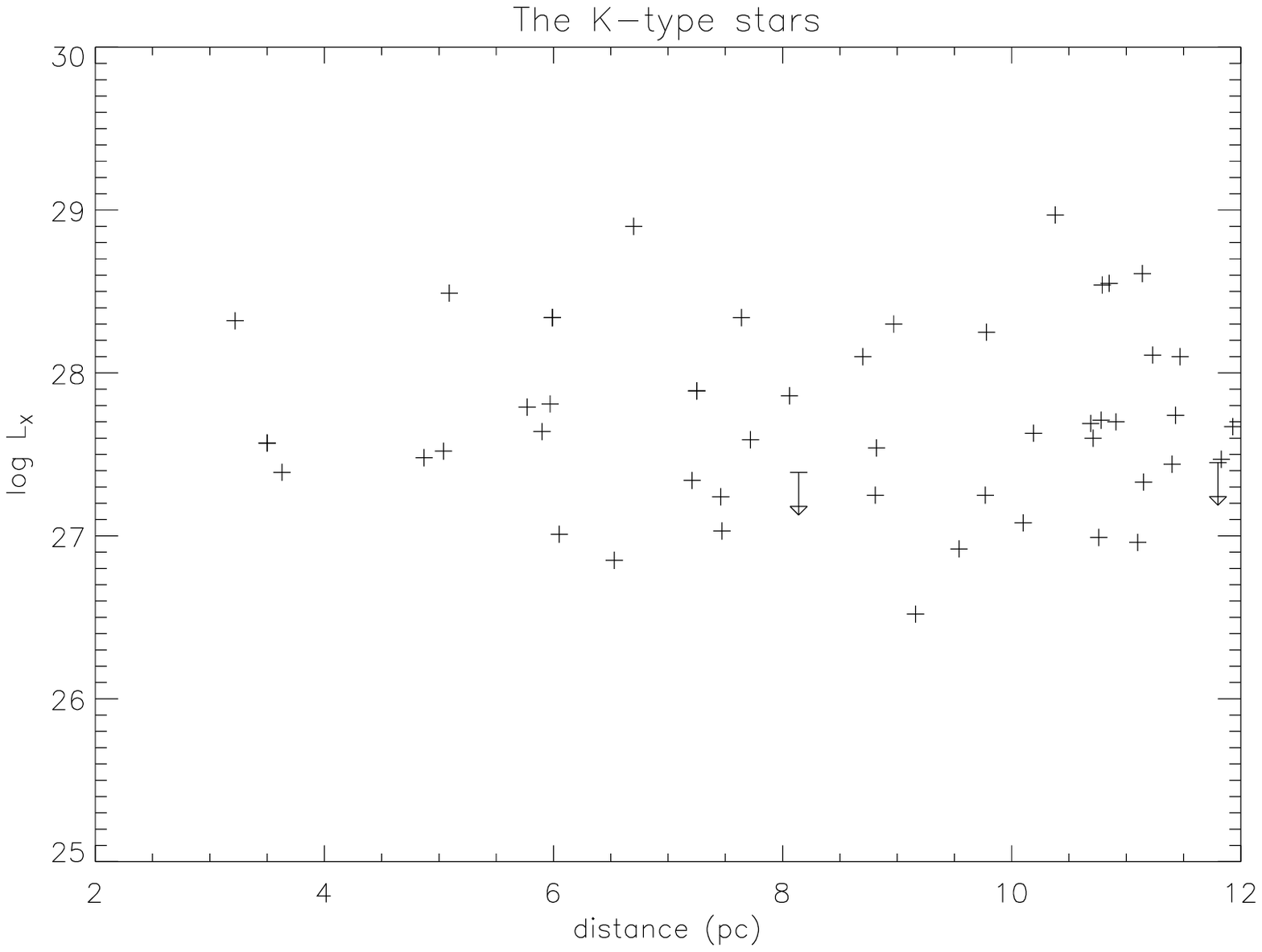}}
\caption[]{\label{K} $L_X$ vs. distance for the K star sample}
\end{figure}
\begin{figure}
\resizebox{\hsize}{!}{\includegraphics{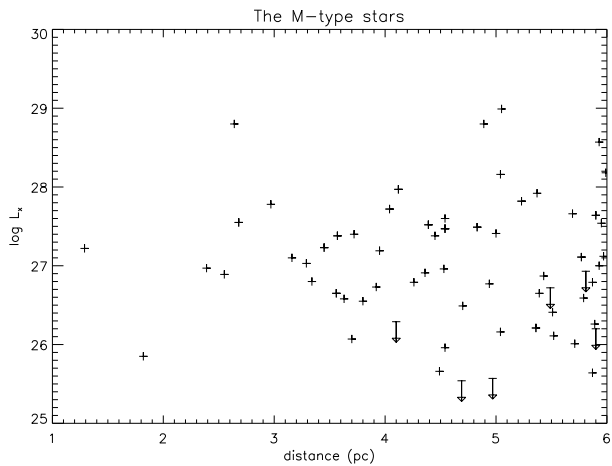}}
\caption[]{\label{M} $L_X$ vs. distance for the M star sample}
\end{figure}

In Fig. \ref{FG}-\ref{M} we plot for the F/G-types stars, the K-type stars and the M-type stars
the X-ray luminosity (in erg/sec) as a function of distance (in pc).
Inspection of Fig. \ref{FG} shows that among 69 F/G stars within a distance of 14 pc around the
Sun, only seven (i.e., the stars Gl~67A, Gl~138, Gl~197, Gl~324A, Gl~354A, Gl~454 and Gl~541A) 
remained undetected. Note that in order
to avoid multiple marks for a star observed more than once, only the observation with the 
longest exposure time was used to create the figures. The 
detection rate within the volume out to 14 pc is therefore 94 \%, and all stars within 
12 pc have been detected. 

The K stars are plotted in Fig. \ref{K}; only two stars (i.e, Gl~776.3 and Gl~884)
out of 51 have not been detected  and hence the detection rate is 96 \%.
We remark in this context that the automatic analysis failed to detect the star
Gl~653 in an HRI pointing.  We therefore retrieved the original X-ray data and
generated an X-ray image having screened the photons to the pulse height
range between channels 2-10. Inspection of this image shows a rather weak source within 
less than two arcsec of the expected proper motioned star position.   Within a 
square box of 10 arcsec by 10 arcsec  12 counts were recorded with 6.1 being 
attributed to background; the remaining counts were then attributed to Gl~653.  Obviously,
a confirmation of this detection is highly desirable.

The M stars are plotted in Fig. \ref{M}; out to a distance of 6 pc 6 stars out of 65
have not been detected; the non-detected stars are Gl~570D, Gl~693, GJ 1002, LP 816-60,
LP 944-20 and DENIS 1048-39.  Gl~570D is a brown dwarf not detected in an HRI pointing of
the Gl~570 system; LP 944-20 is another brown dwarf undetected in a series of HRI
pointings  \citep{Neuhaeuser99}, but 
with a flare detected in a {\it Chandra}
observation \citep{Rutledge00}. Excluding those brown dwarfs all the derived upper limits
except for the star GJ~1002 are derived from survey observations with correspondingly
relatively low sensitivity.
There is no reason to expect that these stars will not be detected once more sensitive
X-ray observations are available. In other words, the formation of X-ray emitting coronae
appears to be universal for late-type main sequence stars.

\subsection {Comparison to previous work}

The most complete previous compilations of X-ray data on nearby late-type stars are due to 
\citet{Schmitt94} for M and K-type stars (Table 1) and to \citet{Schmitt97} for F and G-type 
stars (Tab. 2). Compared to Tab. 1 presented by Schmitt (1997) we first note that now
the HIPPARCOS distance scale has been used.  As a result of this some stars previously
thought to be located within 13 pc are no longer within that distance limit.  Specifically,
the stars Gl~55, Gl~95, Gl~97, Gl~327, Gl~364, Gl~512.1, Gl~534.1A, Gl~611A, Gl~691 and Gl~805 
had to be removed from the sample.  Further, the binary system Gl~107AB has been
observed with the HRI and the individual components are now separated; note that
Gl~107B is of type M and is not included in our nearby sample because of its distance. 
Also, instead of upper limits we now have detections (with the ROSAT HRI) for the
G-type stars Gl~53A and Gl~442A. The M star Gl~53B has also been detected but is again not member
of our nearby stellar sample because of distance, while Gl~442B (= VB 5) at a distance
of 9.24 pc has not been detected in a 10 ksec ROSAT HRI pointing.  \citet{Schmitt95} had
complete K star detections only out to 7 pc, now the limit is 12 pc; on the other hand,
the distance limit for the M-type stars had to be reduced to 6 pc.
Including the stars between 6 and 7 pc would add another 27 stars (from which, however, 7 
remain undetected) 
and therefore significantly enlarge the number of undetected M-stars.

\subsection {X-ray luminosity}

\begin{figure}
\resizebox{\hsize}{!}{\includegraphics{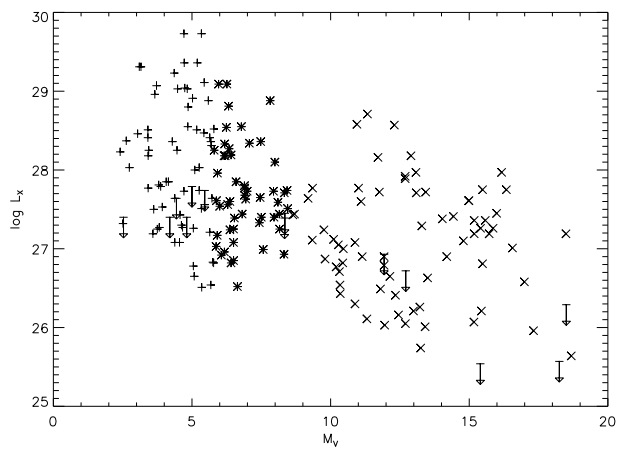}}
\caption[]{\label{MvLx} X-ray luminosity vs. absolute magnitude $M_V$ for the
nearby main-sequence stars; plusses denote F/G-type stars, asterisks K-type stars
and crosses M-type stars.  Note the absence of solar-like stars with X-ray luminosities
of 10$^{26}$ erg/sec and below.
}
\end{figure}

In order to provide an overview over the X-ray luminosities of our sample stars
we show in Fig. \ref{MvLx} the measured X-ray luminosity as a function of
absolute magnitude $M_V$.
Note that the {\bf range} in X-ray luminosity for this 
sample is about three orders of magnitude essentially independent of spectral type,
while the median X-ray luminosity is decreasing with increasing $M_V$.
No objects are found in the right upper corner (there are no 
super-saturation M dwarfs)
nor in the lower left corner, i.e., there are no X-ray dark solar-like stars.
The dependence
of median X-ray luminosity on spectral type suggests to take out the scale effect
introduced by stars of different size by considering instead the mean X-ray surface
flux $F_X$ obtained by dividing $L_X$ by the stellar surface as in \citet{Schmitt97}.

\begin{figure}
\resizebox{\hsize}{!}{\includegraphics{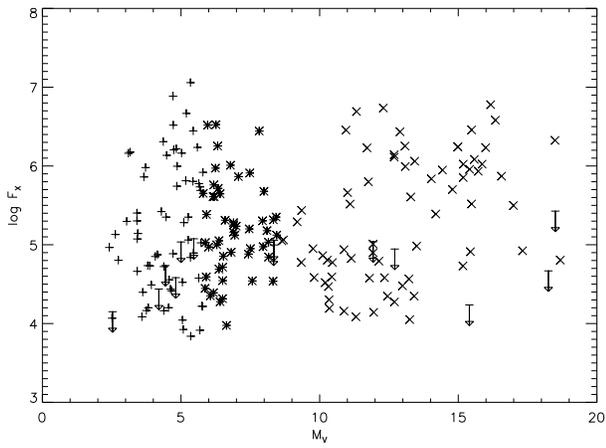}}
\caption[]{\label{MvFx} Mean X-ray surface flux vs. absolute magnitude $M_V$ for the
nearby main-sequence stars; plusses denote F/G-type stars, asterisks K-type stars
and crosses M-type stars.  Note the lower limit of about 10$^4$ erg/cm$^2$/sec of
observed X-ray surface flux. 
}
\end{figure}

\begin{figure}
\resizebox{\hsize}{!}{\includegraphics{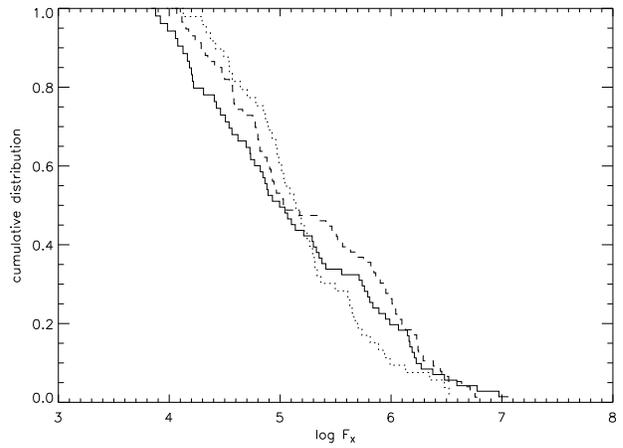}}
\caption[]{\label{cumulative} Cumulative distribution function of the 
mean X-ray surface fluxes for
F/G-type stars (stepped curve), K-type stars (dotted stepped curve) and M-type stars
(dashed stepped curve).  Note the close resemblance of the curves with indistinguishable 
mean and minimum.
}
\end{figure}

A mean quantity such as $F_X$ is little meaningful for ``local'' structures such 
as individual loops
or active regions, however, for ``global'' structures such as coronal holes, it is in fact
a useful and even basic quantity.  The plot of $F_X$ vs. $M_V$ for our sample stars
is shown in Fig.\ref{MvFx}; as in
Fig.\ref{MvLx}, the three different types of symbols represent the F/G-type stars
(plusses), K-type stars (asterisks) and M-type stars (crosses).   
Fig.  \ref{MvFx}
indicates that the mean surface flux distributions of the three classes of stars
show little difference. Possibly the mean surface flux distribution of the very faintest
stars is somewhat larger on average, there is some hint that for stars with
$M_V <$ 13 the lower envelope on $F_X$ may actually increase.  Also note that the K-star
region now appears well filled, while there was a ``gap'' because of lacking sample size in
the previously published data (cf., Fig. 8 in \citet{Schmitt97}).
The formal cumulative distribution functions
are shown in Fig. \ref{cumulative}, showing that indeed the means and lower envelopes of the
three distribution functions agree with each other with statistical accuracy.  
Carrying out formal testing with Smirnov
test shows that the null hypothesis that all three classes of stars are characterized by the
same mean X-ray surface flux distribution function cannot be rejected and is hence
consistent with the data.  We therefore conclude that not only do all cool (dwarf)
stars have coronae, but that also
these coronae have mean X-ray fluxes of at least 10$^4$ erg/cm$^2$/sec.
We mention that these limits are grossly violated (a) by A-type stars, and (b) giants.
For example, for Vega (R = 2.5 $R_{\odot}$) one would compute 
$L_X$ = 4 $\times$ 10$^{27}$ erg/sec; interpreting the upper limit of 1.2 10$^{-3}$ cts/sec,
that \citet{Schmitt97} attributes to UV contamination, as X-ray flux, one obtains
$L_X$ $\sim$ 5.5 10$^{25}$ erg/sec, i.e., two orders of magnitude lower than ``expected'',
and for Arcturus (R = 25 $R_{\odot}$) one obtains $L_X$ $\sim$ 4 $\times$ 10$^{29}$ erg/sec, four 
orders of magnitude higher than the observed upper limit of 3 10$^{25}$ erg/sec \citep{Ayres91}.
The physics of coronal formation in those stars must be very different and in all likelihood those
two stars are devoid of any corona.
 
\subsection {Time variability}

\subsubsection {Variability between survey and pointing data}

\begin{figure}
\resizebox{\hsize}{!}{\includegraphics{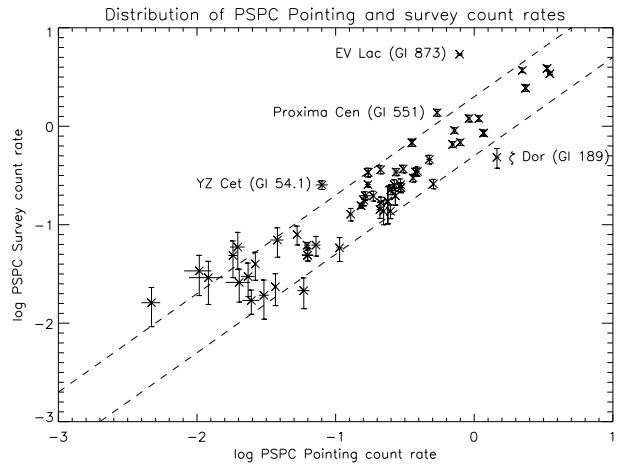}}
\caption[]{\label{surveystat} Comparison of X-ray count rates for sample stars detected both 
in the all-sky survey and the PSPC pointing program; dashed lines indicate a factor 2 
variation from
unity and stars with large deviations from regression line are identified.}
\end{figure}

For a rather large sample of stars multiple X-ray detections are available from the RASS
data and the ROSAT PSPC pointing program, which are all listed in the NEXXUS data base.
  In Fig. \ref{surveystat} we plot the PSPC count rate
observed during the all-sky survey vs. the PSPC count rate observed during 
the pointing program in a double-logarithmic representation for those of our sample stars 
detected in both observing modes; the two long-dashed lines
delineate a factor of two deviation above and below unity.  As is obvious from Fig. \ref{surveystat},
except for four stars, all measurements lie within those dashed lines.  
The time span between survey 
and pointed observations varies considerably from star to star, but is typically of the order
of 1-2 years.  Obviously, variations by more than a factor of 4 are unusual at least on that 
time scale.  The stars EV Lac,
Proxima Cen and YZ Cet, all of which are well known flare stars, have significantly larger 
survey count rates.  In the case of EV Lac this is due to a major flare which occurred during
the survey observations \citep{Schmitt94}, the RASS light curve of Proxima Cen is
discussed by \citet{Fleming93}, the RASS data on YZ Cet is discussed by \citet{Tsik2000}.
In the case of $\zeta$ Dor (spectral type F7) the count rate observed 
in the pointing program was significantly higher than during the survey observations; this applies to both the
PSPC observations with and without filter which had been executed immediately after each other.

\begin{figure}
\resizebox{\hsize}{!}{\includegraphics{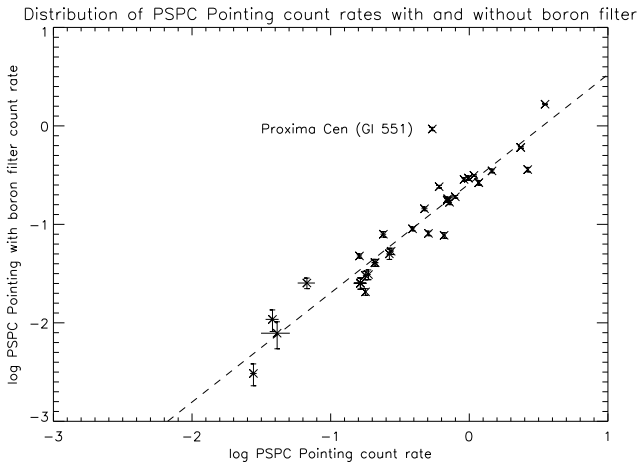}}
\caption[]{\label{filterstat} Comparison of X-ray count rates for all NEXXUS stars detected in the PSPC pointing program with and without boron filter; stars with large deviations from regression line are identified.}
\end{figure}

In Fig. \ref{filterstat} we plot the PSPC pointing mode count rates measured with and 
without the boron filter
for those NEXXUS stars observed with these instrumental setups.  Observations with this setup were
typically (albeit not always) taken immediately adjacent to each other, so that the
time span between the data sets for a given stars should usually be of the order of a few hours.
Of course, the spectral response of the PSPC detector with and without the boron
filter differs especially at soft X-ray energies below the carbon edge, yet, one observes
a well defined regression curve extending over two orders of magnitude.  The only star far away 
from the regression line is again the well known flare star Proxima Cen, which clearly flared during
the boron filter observation.

\begin{figure}
\resizebox{\hsize}{!}{\includegraphics{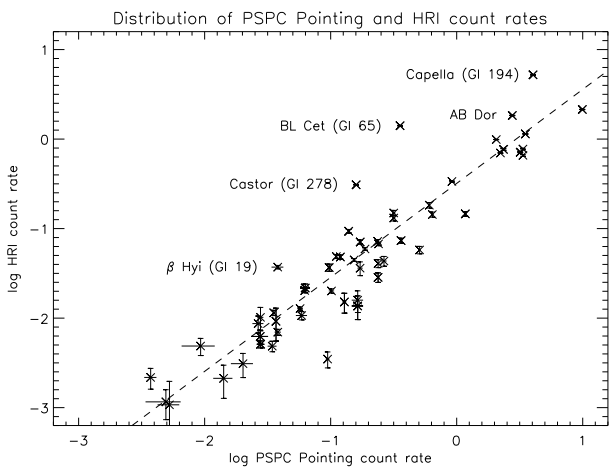}}
\caption[]{\label{hristat} Comparison of X-ray count rates for all NEXXUS stars detected both 
in the PSPC and HRI pointing program ; stars with large deviations from regression line are identified.}
\end{figure}

In Fig. \ref{hristat} we plot the PSPC count rate
measured in pointing program data vs. the HRI count rate in a double-logarithmic 
representation for those of the NEXXUS stars 
detected in both observing modes.  In contrast to the boron filter data, the time span 
between PSPC and HRI pointed observations is much longer and is typically of the order of a few 
years for most stars.  Also note that the spectral response of the HRI differs somewhat from that
of the PSPC. Again, we see a good correlation between PSPC and HRI count rates over two orders
of magnitude with a larger dispersion than observed for the comparison between boron and open 
PSPC data, but with similar dispersion as observed for the pointed and survey PSPC data for the
same stars.
Again, for multiply observed stars only the observation with the longest exposure time was taken into account to compile Fig. \ref{filterstat} and \ref{hristat}.

\subsubsection {Variability of individual stars}

For a small number of stars repeated observations were carried out with the
same instrumental setup. These datasets are discussed for the individual sources in the
following section. 

\begin{figure}
\resizebox{\hsize}{!}{\includegraphics{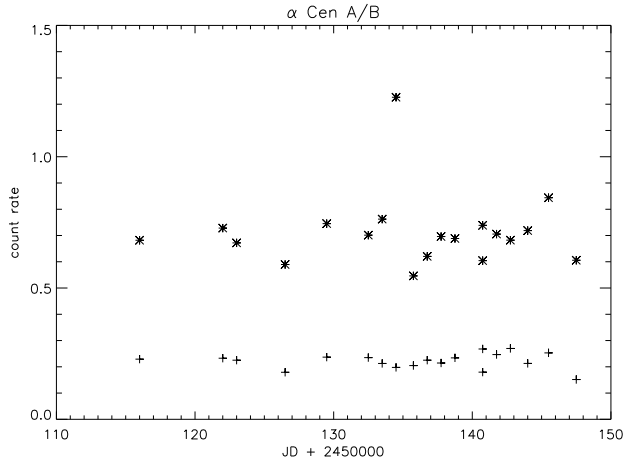}}
\caption[]{\label{lc1} X-ray light curve for $\alpha$ Cen A/B (Gl~559AB) in February 1996; 
asterisks denote measurements for $\alpha$ Cen B, plusses measurements for $\alpha$ Cen A.}
\end{figure}

\begin{figure}
\resizebox{\hsize}{!}{\includegraphics{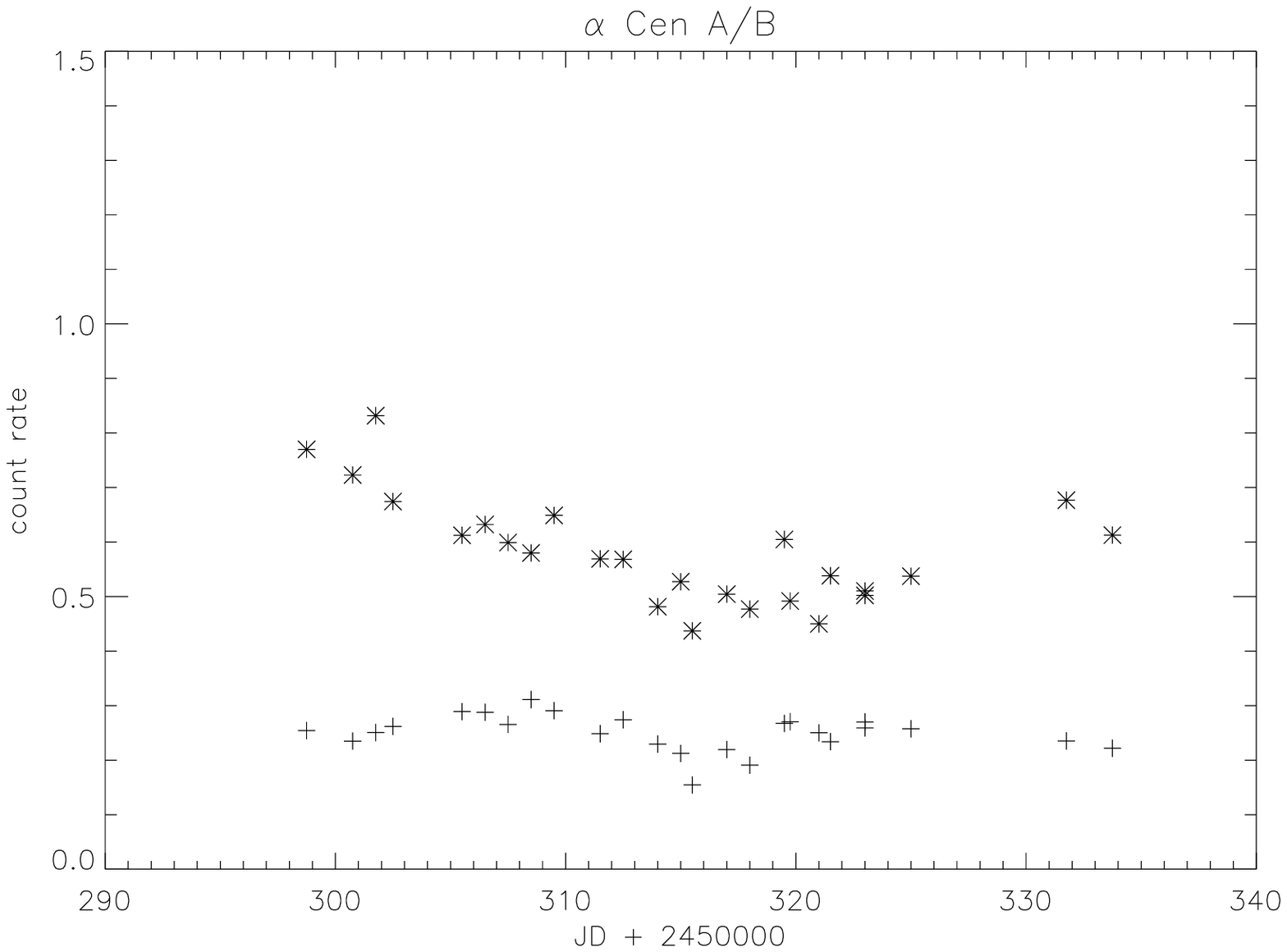}}
\caption[]{\label{lc2} X-ray light curve for $\alpha$ Cen A/B (Gl~559AB) in August 1996;
asterisks denote measurements for $\alpha$ Cen B, plusses measurements for $\alpha$ Cen A.}
\end{figure}

{\it $\alpha$ Cen A/B:}
The $\alpha$ Cen A/B system was observed twice with the ROSAT HRI for an extended 
period of time.  The first set of observations was carried out in February 1996, the second one
in August 1996. The obtained ROSAT HRI light curves for $\alpha$ Cen A/B are shown in Fig. \ref{lc1}
and \ref{lc2}.  In both light curves the stars denote the measurements of  $\alpha$ Cen B,
the crosses those of $\alpha$ Cen A.  As is apparent from Fig. \ref{lc1} and \ref{lc2}, the B
component is brighter than the A component at least every time the system was looked at with 
the HRI.  $\alpha$ Cen A shows small daily variations, the overall level of X-ray emission
was slightly higher during the August observations. $\alpha$ Cen B showed considerably more
X-ray variability than the A  component. In February 1996 the overall emission level was
higher, an excursion to more than 1 HRI cts/sec was observed on day.
During the second observation in August 1996 the count rate 
decreased by about 30 percent over a 20 day interval, afterwards the count rate started 
increasing again.  We interpret the high count rate episode in $\alpha$ Cen B
as a flare, the 20 day decrease in count rate might be due to rotational modulation. 
NEXXUS lists two PSPC pointings on $\alpha$ Cen A/B, one in September 1992 for 3260 seconds
and one in September 1993 for 357 seconds.  During the September 1993 pointing the count rate
was more than twice the count rate observed during the September 1992 and the RASS observations
and the hardness ratio was significantly enhanced.  An inspection of arrival positions of
the individual photons suggests the K star as the source of the increased X-ray radiation in
line with the light curve Fig. \ref{lc1}.  We therefore conclude that $\alpha$ Cen B is a flare 
star.

\begin{figure}
\resizebox{\hsize}{!}{\includegraphics{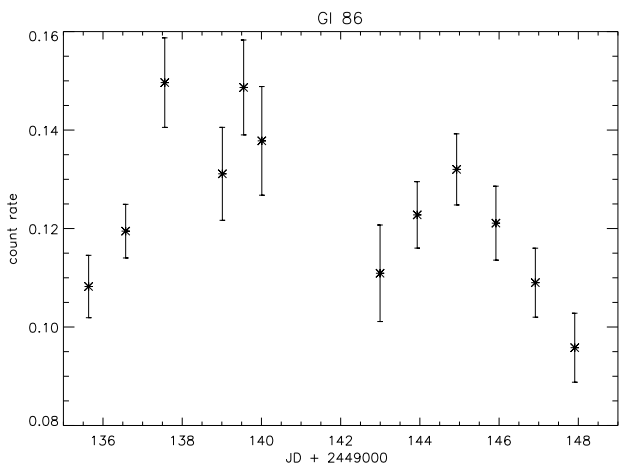}}
\caption[]{\label{f3} X-ray light curve for Gl~86 in May/June 1993}
\end{figure}

{\it Gl~86:}
The star Gl~86 was serendipitously observed during a monitoring campaign on the blazar PKS 0208-512 
with the ROSAT PSPC. 12 pointings almost more or less equally spaced in time were carried out over
a 13 day interval.  The resulting 
ROSAT PSPC light curve for Gl~86 is shown in Fig. \ref{f3}. 
The count rate first increased by almost a factor of 1.5 during the first 
half of the observations. After a 3-day gap the count rate started to increase in the second 
half again from the level it started in the first half, and then it decreased below that level. 
Variations in the X-ray output of Gl~86 are quite apparent; interestingly,
\citet{Marino02} do not find variability in Gl~86 at a confidence level higher than 90 \% 
on short time scales.

{\it Gl~820 A/B:}
The visual binary 61 Cyg A and B was angularly resolved and
extensively observed with the ROSAT HRI to monitor the
long-term X-ray light curve of the two stars.  Since these stars were also simultaneously
observed in Ca H and K,
these synoptic data is presented separately by Hempelmann et al. (2003, in preparation).

{\it LHS~288:}
A special note is appropriate for the star LHS~288.  
Schmitt et al. (1995) report an upper limit of $<$ 0.01 cts/sec at the position of 
LHS 288 from the ROSAT all-sky survey data, while \citet{Marino00} 
report a detection of LHS~288 from a PSPC pointed observation.  
The count rate level of 0.19 cts/sec reported by \citet{Marino00} 
would have been easily detectable in the RASS data, however, the position of the
X-ray source reported by \citet{Marino00} does not agree with expected optical 
position of LHS 288 in CNS4 or in Luyten's catalog.  Investigating this discrepancy we
realized that \citet{Bakos02} were unable to find and confirm even the existence of LHS 288
in their systematic attempt to follow-up high proper motion stars listed in ths LHS catalog.
We therefore decided to investigate the DSS images in the R and V bands in the vicinity 
of the expected position of LHS 288.  We compared the DSS images taken at epochs
1987.0512 and 1991.1096, searched for moving objects and found evidence only 
for one object with significant proper motion, which turned out to be also very red.  
The 1987.0512 position of this object is R.A. 10:44:22.11, $\delta$ -61:12:55.2
(w.r.t. equinox 2000).  The position of the X-ray source reported by \citet{Marino02} is
R.A. 10:44:21.9 and  $\delta$ -61:12:44 at epoch 1993.5575 (w.r.t. equinox 2000).  
According to CNS4, the proper motion of LHS 288 is 1.66 arcsec/year almost due 
north.  Applying the proper motion of LHS 288 as listed in CNS4 to the position
derived for the red moving DSS object results in almost perfect agreement with the 
PSPC X-ray position.   
We therefore conclude that the red moving DSS object is the X-ray source reported by
\citet{Marino00}, and that this object is identical with LHS~288. Thus, LHS~288 has been
re-found and does exist.  We emphasize that also at the new position
no X-ray source can be found in the RASS data with an upper limit of $<$ 0.03 cts/sec.
  
In August 1996 a 12927 sec pointed observation was carried
out on LHS~288 with the ROSAT HRI.   Although the automatic analysis did not show 
any X-ray source at the correct position of LHS~288 and consequently an appropriate entry 
in the ROSAT source catalogs is missing,
a visual inspection of the ROSAT HRI image and a re-analysis of the individual X-ray
photons revealed the presence of a weak, but still significant source at the 
position R.A. 10:44:20.99, $\delta$ -61:12:38.3 in perfect agreement with the previously derived 
(and proper-motioned) PSPC and DSS positions.  The observed count rate of LHS~288 as 
observed in August 1996 was $\approx$ 7.9 10$^{-4}$ HRI cts/sec; applying a factor of
4 to convert to an equivalent PSPC count rate (cf., Fig. \ref{hristat}), we 
find an equivalent PSPC rate of $\approx$ 3.2 10$^{-3}$ PSPC cts/sec, which is 
clearly consistent with the survey non-detection of LHS~288, but not with the
detection of LHS~288 in the pointing program. 
We thus conclude that during the short PSPC pointing on July 21, 1993 the X-ray flux of
LHS~288 was almost two orders magnitude larger than during the RASS observations and during the
HRI pointing in August 1996, and that the PSPC observations on July 21, 1993 caught LHS~288 in an 
unusual state.

\section{Conclusions}

This paper attempts to provide a final and hopefully definitive summary of the ROSAT observations
of nearby stars.  It extends our previous studies of the subject and the now presented source tables
supersede those published previously.  One of our main results is the universality of X-ray emission
among late-type stars with outer convection zones.  As to  the F/G-type stars, all stars at 
distances below 12 pc have been detected, all the upper limits obtained in the distance 
range 12 - 14 pc result from the lower sensitivity survey data.  As to the K-type stars, all 
stars at distances below 8 pc have been detected and only two stars in the distance range 8 - 12 pc
remain undetected; two of those upper limits come from survey data, the other one is derived from
a rather short ROSAT HRI pointing.  As to  the M-type stars, only two stars in the volume out 
to 6 pc 
remain undetected if very late-type stars and brown dwarfs are excluded from consideration; 
one upper 
limit comes from survey data, one from a short PSPC pointing.  If all known stars within 6 pc are
included, the number of non-detected stars increases to 6; in one case (LP~944-20) a Chandra detection
of such an object during a flare
has been obtained.  We therefore conclude that all main-sequence stars with
outer convection zones are surrounded by hot coronae and that the reason for our not being able to 
detect all stars is lack of sufficient sensitivity rather than the intrinsic absence of X-ray
emission.  X-ray dark cool stars on the main sequence do not exist. 

The X-ray luminosities of cool dwarf stars extend over three orders of magnitude with the mean
values decreasing with decreasing spectral type.  Scaling the X-ray luminosity with the stellar
surface results in an activity measure $F_X$ independent of spectral type with essentially
indistinguishable distribution functions.  In particular, stars
with the lowest degree of activity in any subclass always have mean X-ray surface fluxes of
$\approx$ 10$^4$ erg/cm$^2$/sec. \citet{Schmitt97} demonstrated that this flux level is 
approximately the one attained by solar coronal holes,  the much larger sample available now
confirms this conclusion.  Of course, there is no direct proof that the X-ray emission 
originates from magnetically open (rather than closed)
regions in those stars.  However, the fact that the lower surface flux level
is so similar from F-type stars through to M-type stars as well as the lack of observed variability
suggest a global rather than a local
property as cause of the observed similarities.  As an interesting aside we note
that for a star like the Sun the X-ray luminosity contained in 
this ``coronal hole''-component amounts to approximately
6 $\times$ 10$^{26}$ erg/sec formally in the 0.1 - 2 keV, with actually almost all flux 
contained in the 0.1 - 0.5 keV band because of the low X-ray temperatures 
(cf., Fig. 7 in \citet{Schmitt97}).  Relating this value to the bolometric luminosity via
$L_{X,CH}$/L$_{bol}$  results in a value of approximately 1.5 $\times$ 10$^{-7}$, i.e., 
the same value found to describe
the X-ray emission from early type stars (which is also thought to arise in winds).
It is unclear whether this agreement is simply a numerical coincidence or indicative of a physical
connection.

The question of X-ray emission among the very latest main-sequence stars and brown dwarfs 
remains unclear.  ROSAT observations usually do not have the sensitivity level to reach
the required surface flux levels of 10$^4$ erg/cm$^2$sec at the stellar surface; for a typical
star at the bottom of the main sequence this corresponds
to an X-ray luminosity of 6 $\times$ 10$^{24}$ erg/sec (in the 0.1 - 0.5 keV band) 
which is hard to detect even for stars in the immediate solar vicinity.  More importantly, 
the observed large amplitudes of X-ray
variability as seen for vB10 (=Gl~752B;~\citet{Flem2000}), LHS~2065 (\citep{Schm02}), 
LP~944-20 \citep{Rutledge00} and for LHS~288 (this paper) suggest that the
physics of coronal formation for very low mass stars and brown dwarfs may be different and
X-ray emission may be present only in a transient fashion.

This behavior with large amplitude variability has to be juxtaposed to the
the relatively small degree of X-ray variability found for stars of low activity levels in
extensive ROSAT observations.  Comparison of stars with multiple
PSPC observations (in particular for stars with both survey and pointing detections available) shows
that for most stars the two data sets are within a factor of two from unity; the only stars with
discrepant fluxes are known flare stars, where the observed count rate excursions can be clearly
attributed to flare events.  The Sun is known to vary its X-ray output during a solar cycle by
almost two orders of magnitude \citep{acton96}, however, the observed amplitude of variability 
strongly depends on the spectral band considered.  The YOHKOH data described by \citep{acton96}
refer to a somewhat harder X-ray band than the typical ROSAT data are referred to, and therefore
a solar peak-to-peak variation in the ROSAT band of a factor 100 during a solar cycle seems
too extravagant an expectation.  At any rate, from the ROSAT observations we can state that 
count rate variations usually stay within a factor of 4 with exceptions attributable to flares.
So, either the stellar variability level is lower or the time full time scale of variability
has not been adequately sampled by ROSAT. 

\begin{acknowledgements}
The ROSAT work on nearby stars would have been impossible without the ROSAT
all-sky survey.   We thank H. Jahreiss for making available to us his latest version
of CNS4.
We have made extensive use of the ROSAT Data Archive of the Max-Planck-Institut
f\"ur extraterrestrische Physik  (MPE) at Garching, Germany. Also,
this research has made use of the SIMBAD database, operated at CDS,
Strasbourg, France. We particularly thank Dr. J. Pye for providing us with
the ROSAT-WFC data for inclusion into the NEXXUS data base.  We thank our many colleagues
for numerous discussions on the X-ray properties of nearby stars.

\end{acknowledgements}

\bibliographystyle{aa}
\bibliography{papers}

\onecolumn
\small
\begin{longtable}{lllrrrlrrl}
\caption[]{\label{fgstars} Sample stars with absolute magnitude 
between 3.00 and 5.80 (and some F-type stars brighter than 3.00) and a distance up to 14 pc, 
the F- and G-type stars. Table contains the star's name (col 1), its absolute visual 
magnitude (col 2), and its spectral type (col 3) as in CNS4; the distance (in pc, col 4) 
is computed from the CNS4 parallax.  Mode (col 5) denotes the source of the X-ray data, S denoting
survey data, P pointed PSPC data, and H pointed HRI data.  The countrate (col 6) and
its error (col 7) are quoted in cts/sec, the likelihood or SNR are dimensionless (col 8), the
exposure time (col 9) is in seconds, and the X-ray luminosity (col 10) in cgs-units.}\\
\hline
Catalogue&$M_{V}$&Spectral&Distance&Mode&Count Rate&Error&Likelihood&Exposure&log $L_{x}$\\
Name&&Type&(pc)&&&&or S/N&Time\\
\hline
\endfirsthead
\caption[]{Continued}\\
\hline
Catalogue&M$_{V}$&Spectral&Distance&Mode&Count Rate&Error&Likelihood&Exposure&log $L_{x}$\\
Name&&Type&(pc)&&&&or S/N&Time\\
\hline
\endhead
\hline
\endfoot

Gl 5&5.44& K0 V e&13.70&S&0.686800&0.048430&624.0&327&28.97\\
&&&&P&0.793700&0.017800&3690.0&2575&29.03\\
&&&&F&0.190600&0.005840&8385.0&5806&29.11\\
Gl 17&4.55& F9 V&8.59&P&0.022820&0.003500&75.0&2385&27.08\\
Gl 19&3.42& G2 IV&7.47&S&0.069990&0.023200&15.0&195&27.45\\
&&&&P&0.038040&0.004100&114.0&2668&27.18\\
&&&&F&0.010850&0.002680&26.0&1667&27.34\\
&&&&H&0.037140&0.001630&21.0&14526&27.77\\
Gl 27&5.64& K0 V&11.11&H&0.004537&0.000952&4.9&6405&27.21\\
Gl 34A&4.57& G3 V&5.95&S&0.137900&0.021990&71.0&381&27.54\\
&&&&H&0.026470&0.002070&10.2&6391&27.43\\
Gl 53A&5.78& G5 VI&7.55&H&0.004004&0.000662&6.0&12750&26.82\\
Gl 61&3.44& F8 V&13.47&S&0.134900&0.022790&66.0&354&28.25\\
Gl 67A    & 4.44 & G1.5 V   & 12.64&S&$<$0.03801&&2.1&377&27.64\\
Gl 71&5.68& G8 Vp&3.65&S&0.051750&0.013440&24.0&445&26.69\\
&&&&H&0.009015&0.002360&3.9&2000&26.54\\
Gl 75&5.64& K0 V&9.98&S&0.360200&0.026490&505.0&572&28.41\\
Gl 92A&4.85& G0 V e&10.85&S&0.423400&0.039160&285.0&299&28.55\\
Gl 107A&3.87& F7 V&11.23&S&0.152400&0.018690&140.0&527&28.14\\
&&&&H&0.016830&0.002590&5.5&2850&27.79\\
Gl 111&3.72& F6 V&13.97&S&0.906300&0.059260&559.0&288&29.10\\
&&&&P&0.721600&0.020700&3473.0&1779&29.00\\
&&&&F&0.167500&0.008840&4878.0&2136&29.07\\
Gl 124&3.93& G0 V&10.53&S&0.021440&0.007394&13.0&523&27.23\\
&&&&P&0.058940&0.006000&145.0&1821&27.67\\
&&&&H&0.010690&0.001140&8.8&9309&27.53\\
Gl 136&5.10& G2 V&12.12&S&0.095070&0.020590&35.0&328&28.00\\
Gl 137&5.02& G5 V e&9.16&S&1.201000&0.094630&472.0&286&28.86\\
&&&&P&0.917700&0.024400&3994.0&1594&28.74\\
&&&&F&0.285900&0.011600&4366.0&2144&28.94\\
&&&&H&0.336300&0.005450&52.8&11246&28.91\\
Gl 138    & 4.81 & G1 V     & 12.08&S&$<$0.02417&&0.5&338&27.40\\
Gl 139&5.35& G5 V&6.06&S&0.025950&0.009569&10.0&519&26.84\\
&&&&P&0.020170&0.004010&39.0&1738&26.73\\
&&&&H&0.003099&0.000932&3.6&5913&26.51\\
Gl 147&3.59& F8 V&13.72&P&0.011510&0.002340&15.0&2809&27.19\\
Gl 150&3.75& K0 IVe&9.04&S&0.019220&0.008239&9.0&532&27.05\\
&&&&P&0.030400&0.002490&178.0&6383&27.25\\
Gl 177&4.86& G1 V&13.32&S&0.490200&0.045960&272.0&539&28.80\\
Gl 178&3.66& F6 V&8.03&S&1.957000&0.146700&481.0&392&28.96\\
Gl 189&4.36& F7 V&11.65&S&0.484500&0.109600&26.0&1236&28.67\\
&&&&P&1.459000&0.053800&4208.0&500&29.15\\
&&&&F&0.349300&0.015200&3505.0&1507&29.23\\
Gl 197    & 4.20 & G2 IV-V  & 12.65&S&$<$0.02212&&4.2&456&27.40\\
Gl 211&5.79& K1 V e&12.24&S&0.368900&0.032980&331.0&367&28.60\\
&&&&P&0.307700&0.007600&4435.0&5498&28.52\\
Gl 216A&3.82& F6 V&8.97&S&0.343000&0.028840&330.0&488&28.30\\
&&&&P&0.032120&0.004440&64.0&2016&27.27\\
Gl 222AB&4.74& G0 V&8.66&S&1.942000&0.068150&2940.0&431&29.02\\
&&&&F&0.408600&0.008850&9999.0&5284&29.04\\
&&&&H&0.426400&0.011600&31.9&3119&28.96\\
Gl 231&5.05& G5 V&10.15&S&0.017000&0.004695&21.0&1166&27.10\\
&&&&P&0.024660&0.003440&50.0&2660&27.26\\
Gl 280A&2.63& F5 IV-V&3.50&S&3.655000&0.097970&5080.0&400&28.51\\
&&&&P&2.641000&0.027000&9999.0&3654&28.37\\
&&&&F&0.360200&0.018800&3793.0&1061&28.20\\
Gl 302&5.45& G7.5 V&12.58&S&0.032240&0.011320&13.0&355&27.56\\
Gl 324A   & 5.46 & G8 V     & 12.53&S& $<$0.04913&&3.6&337&27.74\\
Gl 354A&2.53&F6 IV&13.49&S&$<$0.01938&&2.1&397&27.40\\
Gl 356A&5.17& G8 V&11.18&S&0.298200&0.026960&254.0&467&28.43\\
&&&&P&0.362400&&779.0&3899&28.51\\
&&&&H&0.073430&0.004580&11.0&3526&28.42\\
Gl 395&4.29& F8 V&12.85&S&0.195400&0.020980&172.0&572&28.36\\
Gl 423AB&4.72& G0 V e&8.35&S&4.539000&0.157500&3380.0&192&29.36\\
Gl 434&5.43& G8 V e&9.54&S&0.348000&0.037120&194.0&314&28.36\\
&&&&P&0.390100&0.014200&3272.0&1956&28.41\\
&&&&F&0.090090&0.004100&3880.0&5572&28.47\\
Gl 442A&5.05& G5 V&9.24&H&0.002430&0.000621&3.6&10109&26.78\\
Gl 449&3.41& F9 V&10.90&S&0.343700&0.030270&264.0&417&28.47\\
&&&&P&0.377500&0.007230&4123.0&7433&28.51\\
Gl 454    & 5.00 & K0 IV    & 12.91&S&$<$0.05198&&7.9&352&27.79\\
Gl 475&4.66& G0 V&8.37&S&0.023470&0.008381&12.0&509&27.07\\
&&&&P&0.036830&&148.0&20537&27.27\\
Gl 482AB&3.10& F0 V&11.83&S&2.039000&0.121100&888.0&146&29.31\\
Gl 502&4.46& G0 V&9.15&S&0.255000&0.037070&78.0&504&28.19\\
&&&&P&0.294400&0.006210&3909.0&8010&28.25\\
Gl 506&5.08& G6 V&8.53&P&0.014160&0.002460&50.0&3113&26.87\\
&&&&H&0.002125&0.000856&2.8&5419&26.65\\
Gl 534&2.41& G0 IV&11.34&S&0.143800&0.023100&55.0&377&28.12\\
&&&&P&0.185100&0.005610&1770.0&6404&28.23\\
Gl 559A&4.38& G2 V&1.34&S&3.876000&0.211800&589.0&420&27.70\\
&&&&P&3.357000&0.033600&9999.0&3260&27.64\\
&&&&H&0.658200&0.014700&25.0&3038&27.53\\
Gl 559B&5.72& K0 V&1.34&S&3.876000&0.211800&589.0&420&27.70\\
&&&&P&3.357000&0.033600&9999.0&3260&27.64\\
&&&&H&0.773600&0.015900&48.9&3038&27.60\\
Gl 566A&5.59& G8 V e&6.70&S&2.440000&0.183400&416.0&400&28.90\\
&&&&P&2.351000&0.020700&9999.0&5213&28.88\\
&&&&F&0.604100&0.011900&9999.0&4270&28.99\\
&&&&H&0.767300&0.017400&32.7&2495&29.00\\
Gl 567&5.70& K2 V&11.54&S&0.362900&0.034130&245.0&398&28.54\\
&&&&P&0.211600&&1735.0&13998&28.31\\
Gl 575A&4.71& F9 V n&12.76&S&3.414000&0.076010&9070.0&612&29.60\\
&&&&P&3.510000&0.040400&9999.0&2166&29.61\\
&&&&H&1.141000&0.005250&190.6&40853&29.73\\
Gl 575B&5.34&dG2&12.76&S&3.414000&0.076010&9070.0&612&29.60\\
&&&&P&3.510000&0.040400&9999.0&2166&29.61\\
&&&&H&1.141000&0.005250&190.6&40853&29.73\\
Gl 598&4.07& G0 V&11.75&S&0.062080&0.014010&30.0&503&27.79\\
&&&&P&0.072190&0.005470&251.0&2726&27.85\\
Gl 601A&2.40& F2 IV&12.31&S&0.144300&0.024700&58.0&365&28.20\\
Gl 603&3.62& F6 V&11.12&S&0.035450&0.011590&10.0&560&27.50\\
Gl 620.1A&4.83& G3/5 V&12.87&S&0.654400&0.044910&646.0&349&28.89\\
&&&&P&0.699200&0.020600&3231.0&1724&28.92\\
&&&&F&0.181900&0.007560&4302.0&3269&29.03\\
Gl 624&4.48& G0 V&12.11&S&1.027000&0.080740&493.0&166&29.03\\
Gl 635AB&2.74&G0 IV&10.80&S&0.127800&0.018560&77.0&494&28.03\\
&&&&P&0.128600&0.004240&2371.0&7752&28.03\\
&&&&H&0.015160&0.003790&3.9&1219&27.71\\
Gl 666A&5.75& G8 V&8.79&S&0.029000&0.013510&8.0&251&27.21\\
&&&&P&0.012090&0.003340&10.0&1250&26.83\\
Gl 695Aa&3.80& G5 IV&8.40&S&0.197000&0.018860&232.0&718&28.00\\
&&&&P&0.163900&0.017000&95.0&587&27.92\\
&&&&F&0.025270&0.003300&98.0&2473&27.81\\
&&&&H&0.015870&0.004320&3.6&910&27.51\\
Gl 702A&5.68& K0 V e&5.09&S&1.649000&0.127200&410.0&347&28.49\\
\nopagebreak
&&&&F&0.247300&0.011500&4242.0&1903&28.36\\
\nopagebreak
&&&&H&0.210900&0.016400&11.0&778&28.20\\
Gl 713A&4.15& F7 V&8.06&S&0.155800&0.009018&506.0&2629&27.86\\
&&&&P&0.152800&0.006840&2255.0&3490&27.85\\
Gl 722&5.26& G5 V&12.98&S&0.045080&0.016000&15.0&283&27.74\\
Gl 771A&3.04& G8 IV&13.71&S&0.247900&0.029290&202.0&343&28.52\\
&&&&H&0.052870&0.006060&7.6&1474&28.46\\
Gl 780&4.62& G8 V&6.11&S&0.073070&0.027780&10.0&124&27.29\\
&&&&H&0.018620&0.002170&8.2&4350&27.30\\
Gl 827&4.38& F8 V&9.22&P&0.009295&0.002730&22.0&1820&26.75\\
&&&&H&0.004878&0.001060&4.9&5914&27.08\\
Gl 848AB&3.41& F5 V&11.76&S&0.259900&0.024040&261.0&505&28.41\\
Gl 853A&4.71& G1 V&13.61&S&0.040330&0.013370&12.0&359&27.73\\
Gl 903&2.51& K1 IVe&13.79&S&0.014180&0.004882&12.0&977&27.29\\
&&&&P&0.015300&0.001930&110.0&5669&27.32\\
Gl 904&3.42& F7 V&13.79&S&0.111500&0.019300&73.0&373&28.18\\
Gl 914A&5.34& G3 V&12.40&S&0.029120&0.011240&12.0&369&27.51\\
\hline
\end{longtable}


\begin{longtable}{lllrrrlrrl}
\caption[]{\label{kstars} Sample stars with absolute magnitude between 5.80 and 8.50 and a distance up to 12 pc, the K-type stars; columns as in Tab. \ref{fgstars}.}\\
\hline
Catalogue&M$_{V}$&Spectral&Distance&Mode&Count Rate&Error&Likelihood&Exposure&log L$_{x}$\\
Name&&Type&(pc)&&&&or S/N&Time\\
\hline
\endfirsthead
\caption[]{Continued}\\
\hline
Catalogue&M$_{V}$&Spectral&Distance&Mode&Count Rate&Error&Likelihood&Exposure&log L$_{x}$\\
Name&&Type&(pc)&&&&or S/N&Time\\
\hline
\endhead
\hline
\endfoot

Gl 33&6.37& K2 V&7.46&S&0.043170&0.013090&17.0&353&27.24\\
Gl 66AB  & 6.26& K2 V&  8.15&S&0.260478&0.039944&99.0&189&29.09\\
Gl 68&5.87& K1 V&7.47&P&0.026610&0.002680&135.0&4076&27.03\\
&&&&H&0.008686&0.001860&4.4&3055&27.14\\
Gl 86&5.92& K0 V&10.91&S&0.058230&0.015840&24.0&375&27.70\\
&&&&P&0.106900&0.005600&1046.0&3587&27.96\\
Gl 92B&6.78&&10.85&S&0.423400&0.039160&285.0&299&28.55\\
Gl 105A&6.53& K3 V&7.21&S&0.058490&0.019040&19.0&232&27.34\\
&&&&H&0.016400&0.000909&16.4&21213&27.39\\
Gl 117&5.96& K2 V&10.38&S&1.200000&0.077170&883.0&220&28.97\\
&&&&P&1.081000&0.016700&9999.0&3880&28.92\\
&&&&F&0.314700&0.006020&9999.0&8609&29.09\\
Gl 144&6.18& K2 V&3.22&S&2.822000&0.233500&327.0&374&28.32\\
&&&&F&0.532700&0.013400&9999.0&3045&28.30\\
&&&&H&0.712300&0.012400&51.8&4585&28.33\\
Gl 166A&5.91& K1 V e&5.04&S&0.179500&0.024990&93.0&341&27.52\\
&&&&H&0.020360&0.004740&4.2&1022&27.17\\
Gl 169&7.99& K7 V&11.47&S&0.133200&0.019570&116.0&429&28.10\\
Gl 183&6.50& K3 V&8.81&S&0.032200&0.012500&9.0&387&27.25\\
Gl 216B&6.41& K2 V&8.97&S&0.343000&0.028840&330.0&488&28.30\\
&&&&P&0.273300&0.011900&2705.0&2013&28.20\\
&&&&F&0.053350&0.003930&2350.0&3540&28.19\\
Gl 250A&6.89& K3 V&8.70&S&0.229300&0.031610&152.0&269&28.10\\
&&&&H&0.029110&0.002950&9.0&3481&27.80\\
Gl 320&6.32& K1 V&11.14&S&0.460300&0.046610&216.0&505&28.61\\
&&&&P&0.475400&0.023800&1128.0&921&28.63\\
&&&&F&0.144100&0.006420&3700.0&3562&28.81\\
Gl 325A&8.43& K5 V&11.43&S&0.059010&0.014350&25.0&423&27.74\\
Gl 370&7.43& K5 V&11.15&H&0.006027&0.001050&5.2&7168&27.33\\
Gl 380&8.15& K7   V&4.87&S&0.176800&0.021330&153.0&494&27.48\\
&&&&P&0.160100&&584.0&3296&27.44\\
Gl 414A&7.96& K8 V&11.93&S&0.046070&0.015680&10.0&317&27.67\\
&&&&H&0.006124&0.001510&4.6&3517&27.40\\
Gl 432A&6.06& K0 V&9.54&H&0.003167&0.000877&3.8&6478&26.92\\
Gl 451A&6.64& G8 VI&9.16&P&0.005465&0.001690&8.0&2765&26.52\\
Gl 453&6.94& K5 V&10.19&H&0.014320&0.001290&11.2&9483&27.63\\
Gl 488&8.33& M0.5Ve&10.78&S&0.061280&0.020860&12.0&221&27.71\\
Gl 505A&6.34& K1 V&11.23&S&0.142800&0.026870&48.0&279&28.11\\
&&&&P&0.209100&0.011400&1338.0&1656&28.28\\
&&&&F&0.040800&0.003290&315.0&3634&28.27\\
Gl 542&6.29& K3 V&11.83&H&0.009136&0.002100&3.2&2445&27.56\\
Gl 566B&7.82& K4 V e&6.70&S&2.440000&0.183400&416.0&400&28.90\\
&&&&P&2.351000&0.020700&9999.0&5213&28.88\\
&&&&F&0.604100&0.011900&9999.0&4270&28.99\\
&&&&H&0.767300&0.017400&32.7&2495&29.00\\
Gl 570A&6.90& K5 V e&5.90&S&0.173100&0.070810&9.0&44&27.64\\
&&&&P&0.238000&&1305.0&5321&27.77\\
&&&&H&0.040760&0.004130&9.4&2443&27.61\\
Gl 617A&8.45& M0 V e&10.69&S&0.059910&0.007541&91.0&1726&27.69\\
&&&&H&0.009817&0.001580&5.9&4503&27.51\\
Gl 631&5.80& K0 V e&9.78&S&0.257900&0.031920&148.0&603&28.25\\
Gl 638&8.16& K7 V&9.77&S&0.026230&0.009423&11.0&468&27.25\\
Gl 653&7.57& K7 V& 10.76&H&0.002910&&7.4\footnotemark[1]&2213&26.99\\
Gl 663AB&6.16& K1 V e&5.99&S&0.854100&0.057780&664.0&291&28.34\\
&&&&P&1.171000&0.025700&9999.0&1783&28.48\\
&&&&F&0.265100&0.013900&3394.0&1429&28.53\\
&&&&H&0.146100&0.008600&14.2&1973&28.18\\
Gl 664&7.46& K5 V e&5.97&S&0.254600&0.031500&157.0&288&27.81\\
&&&&P&0.265000&0.012800&2360.0&1665&27.83\\
&&&&F&0.050210&0.006150&165.0&1391&27.81\\
&&&&H&0.043430&0.005410&4.9&1972&27.65\\
Gl 667AB&6.99& K3 V&7.25&S&0.205700&0.028810&120.0&307&27.89\\
&&&&H&0.035800&0.004370&7.7&1916&27.73\\
Gl 673&8.10& K7   V&7.72&S&0.090570&0.018050&40.0&445&27.59\\
Gl 688&6.37& K3 V&10.71&S&0.048270&0.013440&19.0&440&27.60\\
Gl 702B&7.48& K5 V e&5.09&S&1.649000&0.127200&410.0&347&28.49\\
&&&&F&0.247300&0.011500&4242.0&1903&28.36\\
&&&&H&0.210900&0.016400&11.0&778&28.20\\
Gl 706&6.17& K2 V&11.10&H&0.002579&0.000815&4.0&6294&26.96\\
Gl 713B&6.59&&8.06&S&0.155800&0.009018&506.0&2629&27.86\\
&&&&P&0.152800&0.006840&2255.0&3490&27.85\\
Gl 764&5.88& K0 V&5.77&S&0.255700&0.014290&694.0&1448&27.79\\
&&&&P&0.171300&0.008160&2348.0&2684&27.61\\
&&&&H&0.036140&0.006330&4.9&923&27.54\\
Gl 773.6  & 8.35 & K5 V     & 11.80&S&$<$0.02880&&1.5&307&27.45\\
Gl 783A&6.41& K3 V&6.05&P&0.027630&0.004230&72.0&1883&26.86\\
&&&&H&0.006246&0.001100&5.4&6530&26.82\\
Gl 785&6.00& K0 V&8.82&S&0.061820&0.026730&10.0&126&27.54\\
Gl 820A&7.50& K5   V&3.50&S&0.421700&0.030030&452.0&543&27.57\\
&&&&F&0.057470&0.003430&1282.0&4934&27.40\\
&&&&H&0.071220&0.003280&24.2&6688&27.40\\
Gl 820B&8.32& K7   V&3.50&S&0.421700&0.030030&452.0&543&27.57\\
&&&&F&0.057470&0.003430&1282.0&4934&27.40\\
&&&&H&0.024390&0.001960&11.4&6688&26.93\\
Gl 845&6.89& K5 V e&3.63&S&0.260400&0.028540&169.0&414&27.39\\
&&&&P&0.507900&0.006080&9999.0&14198&27.68\\
&&&&F&0.081000&0.004290&2792.0&4410&27.58\\
&&&&H&0.057770&0.005510&9.3&1958&27.34\\
Gl 879&7.07& K5 V e&7.64&S&0.526300&0.065160&185.0&150&28.34\\
Gl 884    & 8.34 &K5   V   &  8.14&S&$<$0.11527&&6.3&49&27.39\\
Gl 892&6.50& K3 V&6.53&S&0.023110&0.008761&10.0&497&26.85\\
Gl 902&6.81& K3 V&11.40&S&0.029190&0.010000&12.0&462&27.44\\
Gl 909A&6.24& K3 V&10.79&S&0.419100&0.024660&795.0&747&28.54\\
LTT 14084&6.50& K3 V&10.10&S&0.016520&0.006606&9.0&813&27.08\\
\hline
\end{longtable}


\begin{longtable}{lllrrrlrrl}
\caption[]{\label{mstars} Sample stars with absolute magnitude fainter than 8.50 and a distance up to 6 pc, M-type stars; columns as in Tab. \ref{fgstars}.}\\
\hline
Catalogue&M$_{V}$&Spectral&Distance&Mode&Count Rate&Error&Likelihood&Exposure&log Lx\\
Name&&Type&(pc)&&&&or S/N&Time\\
\hline
\endfirsthead
\caption[]{Continued}\\
\hline
Catalogue&M$_{V}$&Spectral&Distance&Mode&Count Rate&Error&Likelihood&Exposure&log L$_{x}$\\
Name&&Type&(pc)&&&&or S/N&Time\\
\hline
\endhead
\hline
\endfoot

Gl 1&10.35& M3   V&4.36&S&0.059340&0.024300&9.0&156&26.91\\
&&&&P&0.019690&0.002380&104.0&4604&26.43\\
Gl 15A&10.32& M1.5 V&3.57&S&0.260200&0.026180&202.0&453&27.38\\
&&&&H&0.013900&0.003060&4.4&1606&26.71\\
Gl 15B&13.29& M3.5 V&3.57&S&0.260200&0.026180&202.0&453&27.38\\
&&&&H&0.053440&0.005780&8.7&1606&27.29\\
Gl 34B&8.64& K7   V&5.95&S&0.137900&0.021990&71.0&381&27.54\\
&&&&H&0.026470&0.002070&10.2&6391&27.43\\
Gl 54.1&14.19& M4.5 V&3.72&S&0.253800&0.025840&200.0&461&27.40\\
&&&&P&0.079810&0.008050&142.0&1359&26.90\\
Gl 65AB&15.41& M5.5 V&2.68&S&0.680300&0.056820&318.0&484&27.55\\
&&&&P&0.355700&&3253.0&16136&27.26\\
&&&&H&1.410000&0.010100&121.7&13884&28.46\\
Gl 83.1&14.03& M4.5 V&4.45&S&0.167400&0.024950&111.0&295&27.38\\
Gl 166C&12.69& M4.5 V&5.04&S&0.796400&0.051660&572.0&344&28.16\\
&&&&H&0.107100&0.010400&9.8&1022&27.89\\
Gl 169.1A&12.34& M4   V&5.51&P&0.011920&0.002220&41.0&3323&26.41\\
Gl 191&10.88& M1 p V&3.92&S&0.048740&0.019720&10.0&248&26.73\\
&&&&P&0.018180&0.001320&353.0&14268&26.30\\
Gl 205&9.19& M1.5 V&5.69&S&0.196300&0.022530&152.0&455&27.66\\
&&&&P&0.187000&0.006120&2014.0&5156&27.64\\
Gl 213&12.70& M4   V&5.79&S&0.016150&0.006921&9.0&461&26.59\\
&&&&P&0.004701&0.000729&43.0&13362&26.05\\
Gl 229AB&9.34& M1 V&5.77&S&0.053500&0.011580&39.0&592&27.11\\
Gl 234AB&13.08&M4.0 V&4.12&S&0.767000&&&490&27.97\\
Gl 251&11.31& M3   V&5.52&P&0.005924&0.001500&30.0&4048&26.11\\
Gl 273&11.95& M3.5 V&3.80&S&0.034000&0.014910&9.0&247&26.55\\
&&&&P&0.010390&0.002310&37.0&2932&26.03\\
Gl 285&12.30& M4   V&5.93&S&1.467000&0.078880&1220.0&253&28.57\\
Gl 300&13.22& M3.5 V&5.89&P&0.007263&0.001580&28.0&4096&26.26\\
Gl 388&10.95& M3   V&4.89&S&3.701000&0.194400&1500.0&104&28.80\\
&&&&P&2.226000&0.013500&9999.0&25092&28.58\\
&&&&H&0.698700&0.006710&94.6&15302&28.68\\
Gl 406&16.56& M6   V&2.39&S&0.227500&0.025720&183.0&384&26.97\\
&&&&P&0.250000&0.009960&2566.0&2604&27.01\\
Gl 411&10.46& M2   V&2.55&S&0.167000&0.024170&83.0&341&26.89\\
&&&&P&0.212400&0.004760&3844.0&10006&27.00\\
Gl 412A&10.34& M1   V&4.83&S&0.184800&0.026840&111.0&364&27.49\\
&&&&H&0.005208&0.000526&5.0&24751&26.54\\
Gl 412B&15.99& M5.5 V&4.83&S&0.184800&0.026840&111.0&364&27.49\\
&&&&H&0.042180&0.001320&25.4&24751&27.45\\
Gl 445&12.14& M3.5 V&5.39&P&0.021170&0.004150&41.0&1485&26.65\\
Gl 447&13.50& M4   V&3.34&S&0.079610&0.017270&41.0&364&26.80\\
&&&&P&0.052780&&174.0&3885&26.63\\
Gl 473AB&14.99& M5.5 V&4.39&S&0.238600&0.026880&147.0&430&27.52\\
&&&&P&0.295600&0.012900&2437.0&1922&27.61\\
Gl 526&9.80& M1.5 V&5.43&P&0.035310&0.007460&36.0&747&26.87\\
Gl 551&15.48& M5.5&1.29&S&1.374000&0.110700&445.0&374&27.22\\
&&&&P&0.541100&0.005680&9999.0&17157&26.81\\
&&&&F&0.930500&0.035500&9999.0&725&27.74\\
Gl 570BC&9.35& M1   V&5.90&S&0.173100&0.070810&9.0&44&27.64\\
&&&&P&0.238000&&1305.0&5321&27.77\\
&&&&H&0.028530&0.003470&7.4&2443&27.46\\
Gl 570D&&T8&5.90&H&$<$0.00158&&1.3\footnotemark[1]&2430&26.20\\
Gl 588&10.44& M2.5 V&5.93&S&0.039920&0.012640&13.0&416&27.00\\
&&&&P&0.026350&&61.0&4331&26.82\\
Gl 628&11.93& M3   V&4.26&S&0.047450&0.011080&27.0&554&26.79\\
Gl 674&11.10& M3   V&4.54&S&0.266600&0.040960&76.0&193&27.60\\
Gl 682&12.45& M4.5 V&5.04&P&0.007986&0.002780&8.0&1433&26.16\\
Gl 687AB&10.88& M3   V&4.53&S&0.062050&0.004069&374.0&5788&26.96\\
&&&&P&0.062300&0.003340&1259.0&6274&26.96\\
&&&&H&0.020350&0.001200&15.3&15193&27.08\\
Gl 693&11.93&M3.5 V&5.81&S&$<$0.03522&&0.7&129&26.93\\
Gl 699&13.25& M4   V&1.82&S&0.029760&0.011060&13.0&399&25.85\\
&&&&P&0.023370&0.002620&105.0&4127&25.74\\
Gl 725AB&11.14& M3   V&3.56&S&0.049080&0.006292&91.0&1781&26.65\\
&&&&P&0.062860&0.005710&158.0&2263&26.76\\
&&&&H&0.021700&0.002330&8.3&4178&26.90\\
Gl 729&13.09& M3.5 V&2.97&S&0.942000&0.080930&334.0&326&27.78\\
&&&&H&0.204200&0.006880&27.5&4305&27.71\\
Gl 752A&10.28& M3   V&5.87&P&0.035360&&139.0&10492&26.94\\
&&&&H&0.011460&0.000778&14.6&21758&27.05\\
Gl 752B&18.68& M8   V&5.87&H&0.000444&0.000323&3.0&21757&25.64\\
Gl 754&13.41& M4.5&5.71&P&0.004324&0.000974&15.0&6529&26.01\\
Gl 825&8.70& M0.5 V&3.95&S&0.137300&0.022020&83.0&403&27.19\\
&&&&P&0.248600&0.005250&3589.0&9467&27.44\\
Gl 832&10.20& M1 V&4.94&P&0.033740&0.004580&54.0&2041&26.77\\
Gl 860AB&11.76& M3   V&4.04&S&0.448600&0.033550&468.0&427&27.72\\
Gl 866AB&15.18& M5   V&3.45&S&0.196900&0.038900&57.0&161&27.23\\
&&&&P&0.269900&0.004640&3800.0&12892&27.36\\
Gl 873&11.71& M3.5 V&5.05&S&5.384000&0.098370&9999.0&560&28.99\\
&&&&P&0.784500&&5458.0&4463&28.16\\
Gl 876&11.80& M3.5 V&4.70&P&0.019320&0.003770&34.0&1766&26.49\\
Gl 887&9.76& M2   V&3.29&S&0.137400&0.037470&28.0&148&27.03\\
&&&&P&0.222000&0.004350&3425.0&12729&27.24\\
Gl 905&14.79& M5.5 V&3.16&S&0.177200&0.023810&100.0&350&27.10\\
Gl 908&10.11& M1   V&5.97&S&0.051160&0.013450&27.0&414&27.12\\
GJ 1002&15.40&M5.5 V&4.69&P&$<$0.00219&&0.3&1673&25.54\\
GJ 1005AB&12.99& M3.5 V&5.36&P&0.007931&0.002540&10.0&1643&26.21\\
GJ 1061&15.17& M5.5 V&3.70&P&0.011900&0.003690&16.0&1321&26.07\\
GJ 1111&16.99& M6.5 V&3.63&P&0.039730&&264.0&10312&26.58\\
GJ 1116AB&15.48& M5.5 V&5.23&S&0.340000&0.034000&231.0&337&27.82\\
&&&&P&0.171700&&389.0&3021&27.53\\
&&&&H&0.070920&0.003670&17.9&5319&27.75\\
GJ 1245ABC&15.18& M5.5 V&4.54&S&0.198200&0.016990&331.0&797&27.47\\
&&&&H&0.026270&0.003120&6.8&2849&27.19\\
LHS 288&15.61&M5.5&4.49&P&0.197300&0.016300&302&730&27.46\\
&&&&H&0.00079&0.00028&11.7\footnote[1]{Given value is the source likelihood}&12927&25.66\footnote[2]{Source not included in the HRI catalog. See Sec. 5.4.2 for details.}\\
LHS 292&17.32& M6.5 V&4.54&P&0.006161&0.001330&39.0&5146&25.96\\
LP 71-82&14.43&dM7 e&5.00&S&0.143900&0.004378&2290.0&9529&27.41\\
LP 816-60&12.71&M&5.49&S&$<$0.02430&&1.9&326&26.72\\
LP 944-20&18.25&M9  V&4.97&H&$<$0.00415&&&220800&25.57\footnote[3]{X-ray data taken from \citet{Neuhaeuser99}}\\
LTT 17897&12.69& M3.5 V&5.37&S&0.404000&0.044720&193.0&475&27.92\\
L 34-26&12.90&dM5 e&5.99&S&0.594400&0.070090&171.0&753&28.18\\
DENIS 1048-39&18.50&M9  V&4.10&S&$<$0.01599&&0.5&409&26.29\\
\hline
\end{longtable}
\normalsize

\twocolumn

\end{document}